\documentclass[trackchanges,twocolumn]{aastex701} % linenumbers
\usepackage{color}

\newcommand{\gsim}{\hspace{0.3em}\raisebox{0.4ex}{$>$}\hspace{-0.75em}\raisebox{-.7ex}{$\sim$}\hspace{0.3em}}

\begin{document}

%%%%%%%%%%%%%%%%%%%%%%%%%%%%%%%%%%%%%%%%%%%%%%%%%%%%%%%%%%%%%%%%%%%%%%%%%%%%%%%%%%%%%%%%%%%%
% Title, authors and keywords
%%%%%%%%%%%%%%%%%%%%%%%%%%%%%%%%%%%%%%%%%%%%%%%%%%%%%%%%%%%%%%%%%%%%%%%%%%%%%%%%%%%%%%%%%%%%

\title{
The Milky Way Tomography with Subaru Hyper Suprime-Cam:\\
Implications for the past orbit of the Large Magellanic Cloud}

\author[orcid=0009-0009-9769-534X, gname=Yoshihisa, sname='Suzuki']{Yoshihisa Suzuki}
\affiliation{Astronomical Institute, Tohoku University, Aoba-ku, Sendai 980-8578, Japan}
\email[show]{yoshihisa.suzuki@astr.tohoku.ac.jp}  

\author[orcid=/0000-0002-9053-860X, gname=Masashi, sname='Chiba']{Masashi Chiba} 
\affiliation{Astronomical Institute, Tohoku University, Aoba-ku, Sendai 980-8578, Japan}
\email{chiba@astr.tohoku.ac.jp}

\author[orcid=0000-0002-4013-1799, gname=Rosemary, sname='Wyse']{Rosemary F. G. Wyse}
\affiliation{The William H. Miller III, Dept. of Physics \& Astronomy, Johns Hopkins University, Baltimore, MD 21218, USA}
\email{wyse@jhu.edu}

%%%%%%%%%%%%%%%%%%%%%%%%%%%%%%%%%%%%%%%%%%%%%%%%%%%%%%%%%%%%%%%%%%%%%%%%%%%%%%%%%%%%%%%%%%%%
% Abstract
%%%%%%%%%%%%%%%%%%%%%%%%%%%%%%%%%%%%%%%%%%%%%%%%%%%%%%%%%%%%%%%%%%%%%%%%%%%%%%%%%%%%%%%%%%%%

\begin{abstract}
We report the discovery of diffuse stellar substructure in the Milky Way's outer halo toward Bo\"otes, unveiled by deep imaging data of the Subaru/Hyper Suprime-Cam. 
This substructure is detected as an excess of faint main-sequence stars, at heliocentric distances beyond 30~kpc, extending over at least 100~$\mathrm{deg^2}$. 
To infer its origin, we compare the projected spatial distribution of these stars to that of simulated tidal debris from the Large Magellanic Cloud (LMC), under the assumptions that the LMC is on either its first or second passage of the Milky Way. 
We found that the observed overdensity lies in a region of the halo where debris from the LMC is expected if it is on its initial pericentric phase 7-8 Gyr ago, which is predicted in the second-passage model, while the first-passage model is unable to explain the observed substructure.
Chemo-kinematical data are required to further constrain its past orbit and to understand the origin of this new halo substructure, as should be obtained in the near future with deep photometric surveys such as UNIONS and LSST, and wide-field spectroscopy such as possible with PFS and DESI.  
\end{abstract}

%% https://astrothesaurus.org
%% https://astrothesaurus.org/concept-select/
\keywords{\uat{Large Magellanic Cloud}{903} ---  \uat{Milky Way evolution}{1052} --- \uat{Milky Way formation}{1053} --- \uat{Milky Way stellar halo}{1060}}

%%%%%%%%%%%%%%%%%%%%%%%%%%%%%%%%%%%%%%%%%%%%%%%%%%%%%%%%%%%%%%%%%%%%%%%%%%%%%%%%%%%%%%%%%%%%
% Section 1: Introduction
%%%%%%%%%%%%%%%%%%%%%%%%%%%%%%%%%%%%%%%%%%%%%%%%%%%%%%%%%%%%%%%%%%%%%%%%%%%%%%%%%%%%%%%%%%%%

\section{Introduction} 

%%%% Fig.1 %%%%%%%%%%%%%%%%%%%%%%%%%%%%
\begin{figure*}[ht!]
    \plotone{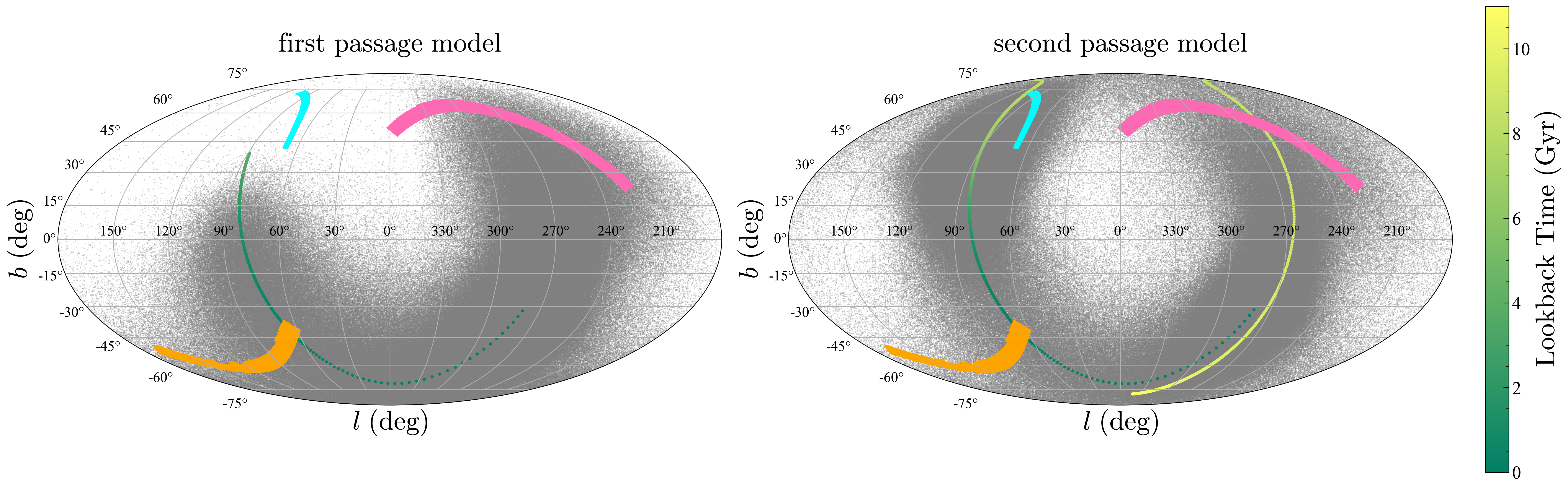}
    \caption{
    The survey footprint in the HSC-SSP and the expected tidal debris from the different orbital model of the LMC. 
    The left panel shows the case for first-passage model of L2M10first from \citet{Vasiliev+2024}. 
    The gray dots shows the expected tidal debris. 
    The sequential colored dots show the orbit of LMC from the past (yellow) to now (green).
    The right panel shows the case for the second-passage model of L2M10 from \citet{Vasiliev+2024}. 
    The gray dots and the sequential color is the same as the left panel. 
    For both panels, the North, Spring, and Fall field in the Subaru HSC-SSP survey are drawn as the cyan, pink, orange area, respectively. 
    This figure indicates that the hints for the past orbit of the LMC can be seen as the substructure toward North field. 
    }
    \label{fig1}
\end{figure*}
%%%%%%%%%%%%%%%%%%%%%%%%%%%%%%%%%%%%%%%

According to the current concordance cosmological model based on $\Lambda$CDM, structures grow through the hierarchical merging of smaller systems. 
On the scales of galaxies such as the Milky Way (MW), numerical simulations have revealed that such processes are encoded in stellar halos \citep[e.g.][]{Johnston+1996, Bullock+2005, Cooper+2010} in the form of persistent features in chemo-dynamical phase-space and streams and shells in coordinate space. 
Thus, the stellar halo retains the richest information about the past merging history of galaxies. 

For our galaxy, the MW, past accretion events have been identified from its inner halo by combining kinematic phase-space information with chemical abundances. 
With this technique, it was found that debris from the so-called Gaia Enceladus/Sausage \citep{Helmi+2018, Belokurov+2018} is the dominant constituent of the inner MW halo \citep{Naidu+2020}, while other kinematic substructures, such as the Helmi stream, are also classified through chemical tagging \citep[e.g.\@, ][]{Horta+2023}.

On the other hand, the Large Magellanic Cloud (LMC) is now thought to play an important role in the formation and evolution of the outer halo of the MW. 
There is strong evidence, for example from its apparent dynamical effect on stellar streams in the halo \citep[e.g.,][]{Erkal+2019}, that the total mass of the LMC is as much as 10\% of that of the MW, making it the most massive satellite galaxy.  
Based on such a massive LMC, \citet{GaravitoCamargo+2019} conducted N-body simulations of the LMC–MW interaction and demonstrated that a collective response is induced in the MW dark matter halo due to the displacement of its center, and that a transient wake arises from dynamical friction exerted by the halo material on the LMC.
Observationally, \citet{Conroy+2021} detected a wide-angle overdensity in star-counts in the outer MW's stellar halo, together with a local overdensity closer to the present location of the LMC, and behind it on its assumed orbit, using K-giants as distance tracers. 
These findings match well with the predicted response of the MW to the infall of the LMC and suggests that the outer halo is indeed in disequilibrium, due to gravitation perturbation by the LMC.

The fact that the LMC has a dynamical effect on the MW is now reasonably well-established, but we still do not accurately understand the past orbit of the LMC.
Since the early attempts of \citet{Lin+1977} and \citet{Murai+1980} to reproduce the prominent feature in neutral hydrogen maps associated with Magellanic Clouds, the so-called Magellanic stream, as representing tidal debris from the most recent passage of the LMC after many orbits, subsequent simulations have been carried out under the assumption of multiple past orbital passages \citep[e.g.\@, ][]{Gardiner+1996,Diaz+2012}. 
However, \citet{Besla+2007} proposed that the LMC is instead approaching the MW for the first time, based on the high value of the three-dimensional velocity (with respect to the MW) they derived by combining the estimated center-of-mass proper motion from Hubble Space Telescope imaging data with the line-of-sight velocity. 
Also, \citet{Conroy+2021} claims that the detection of the local overdensity is independent evidence of the LMC first-passage model.
More recently,  \citet{Sheng+2024} emphasized that the uncertainties in the masses of the MW and LMC will affect determinations of the past orbit of the LMC.
Further, \citet{Vasiliev+2023} and \citet{Vasiliev+2024} demonstrated that neither the transient wake behind the LMC nor the larger-scale collective response should be sensitive to past orbits, indicating that such information alone, referring only to the effect the interaction with the LMC has on the MW, may not constrain the past orbit of the LMC.

In order to distinguish between the first and second passage models of the LMC, \citet{Vasiliev+2024} proposed the necessity to look at the effect on the LMC. 
In particular, \citet{Vasiliev+2024} pointed out that there is a clear difference in the spatial distribution of the expected tidal debris from the LMC under these two different scenarios.
Figure~\ref{fig1} shows the expected tidal debris from the LMC in Galactic coordinates based on the last timestep from simulations taken from publicly available dataset\footnote{https://zenodo.org/records/8015660} from \citet{Vasiliev+2024}, together with the survey footprints of the Subaru/Hyper Suprime-Cam (HSC) that we utilize here, referred to as the Spring (pink), Fall (orange), and North (cyan) fields.
The left panel is the case for the first-passage model of L2M10first from \citet{Vasiliev+2024}, which assumes the mass of the MW of $10\times10^{11}M_{\odot}$ and the mass of the LMC of $2\times10^{11}M_{\odot}$, while the right panel is the case for the second-passage model of L2M10, which assumes the same mass of L2M10first model, but the first passage starting from the apocenter.
Clearly, the tidal debris from the LMC is widely distributed in the second quadrant of the Galactic coordinate only in the second-passage model, and this region overlaps with the North field.
Thus, we expect to obtain new insights into the past orbit of the LMC by exploring structure in the MW's stellar halo toward the North field.
Since the debris is expected to be sparse, we require high-density tracers, such as main-sequence turn off (MSTO) stars.
At this point, deep and wide imaging data obtained with HSC on the Subaru telescope is ideal to detect such low surface brightness structure.

In this letter, we aim to investigate the structure of the MW's stellar halo to test whether or not the possible signal of the past orbit of the LMC exists.
In Section 2, we describe the dataset, taken with the Subaru/HSC, and the methodology we developed to quantify the structure of the stellar halo.
In Section 3, we show our results.
In Section 4, we discuss the possible origins of detected substructure, and summarize our conclusions.
For further details concerning data reduction and analysis, please refer to \citet{Suzuki+2024} (Paper~I) and Suzuki et al. 2026 (Paper~II).

%%%%%%%%%%%%%%%%%%%%%%%%%%%%%%%%%%%%%%%%%%%%%%%%%%%%%%%%%%%%%%%%%%%%%%%%%%%%%%%%%%%%%%%%%%%%
% Section 2: Data and Analysis Method
%%%%%%%%%%%%%%%%%%%%%%%%%%%%%%%%%%%%%%%%%%%%%%%%%%%%%%%%%%%%%%%%%%%%%%%%%%%%%%%%%%%%%%%%%%%%

%%%% Fig.2 %%%%%%%%%%%%%%%%%%%%%%%%%%%%
\begin{figure*}[ht!]
    \plotone{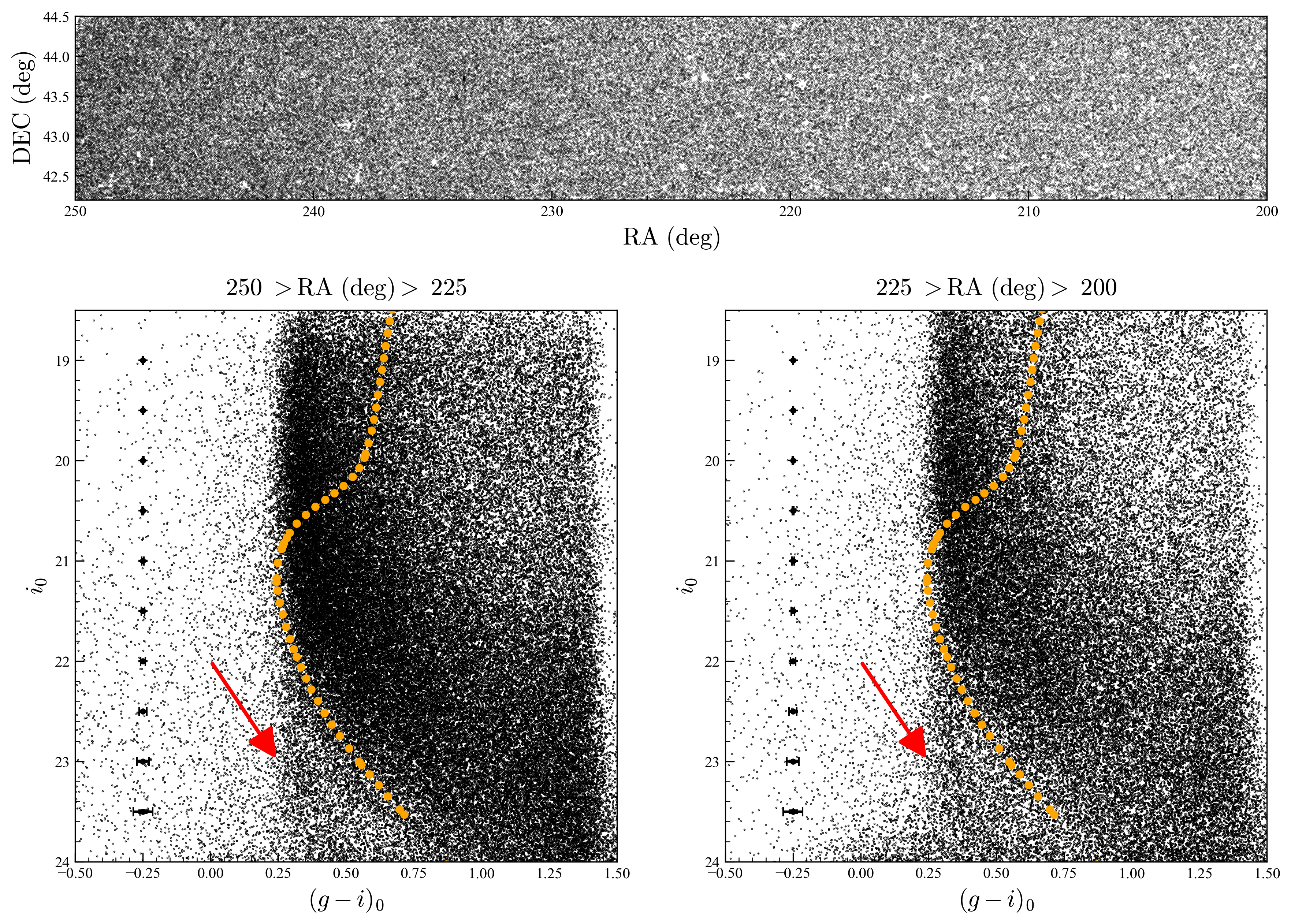}
    \caption{
    Halo stars toward the North field. 
    The upper panel shows the spatial distribution of halo stars in equatorial coordinates, while the two lower panels show the color magnitude diagrams in a fixed range of declination (from 42.5 to 44.5~deg) and two ranges of right ascension (left: from 225 to 250~deg; right: from 200 to 250~deg).   
    The orange dotted line in each panel represents an isochrone with age of 11.5~Gyr, metallicity of $-1.6$, and assumed distance from the Sun of 30~kpc.
    The black dots with error bars show typical photometric uncertainties as a function of $i_{0}$
    This figure indicates that substructure exists in the stellar halo beyond 30~kpc as indicated by the red arrow, and extends over at least 100~$\mathrm{deg}^2$.
    }
    \label{fig2}
\end{figure*}
%%%%%%%%%%%%%%%%%%%%%%%%%%%%%%%%%%%%%%%

\section{Data and Analysis Method}

We utilize the recent photometric catalog (S21A) obtained by the Subaru/HSC Subaru Strategic Program (HSC-SSP) \citep{Aihara+2018a, Aihara+2018b, Aihara+2019, Aihara+2022}. 
This catalog was generated by the hscPipe~8.4, updated with an FGCM code (hscPipe~8.5.3), based on the pipeline developed for the Vera~C.~Rubin Observatory \citep{Ivezic+2008, Juric+2017, Bosch+2019} and calibrated against Pan-STARRS1 photometry and astrometry \citep{Schlafly+2012, Tonry+2012, Magnier+2013, Chambers+2016}. 
This dataset is unique for the wide survey footprint, consisting of North, Spring, and Fall fields (see Figure~\ref{fig1}), covering 1,100~$\mathrm{deg}^2$, and for its deep photometry, reaching down to a 5-sigma limiting magnitude of 26.2 in the $i$-band.
Here, we particularly focus on the North field (cyan in Figure~\ref{fig1}) in the Wide Layer to investigate substructure in the stellar halo that could consist of tidal debris from the LMC that was stripped during its first perigalactic passage, assuming an orbit for the LMC that places it on its second passage (see the right panel of Figure~\ref{fig1}).

We obtain the sample of halo stars by first considering all point sources, as flagged by \texttt{i\_extendedness\_value = 0.0}, then selecting those with $i$-band magnitude brighter than 24.5, fainter than which the probability of a star is below 0.5. 
Further refinement of the sample of halo stars is based on the ($g$-$r$) vs ($r$-$i$) two-color diagram, as shown in Paper~I.

Since the total stellar mass of early tidal debris from the LMC is predicted to be sufficiently low, intrinsically bright evolved stars such as blue horizontal branch stars and K-giants are too rare to trace such structure. 
Thus, we use (old) MSTO stars, which are relatively abundant and for which photometric distance estimates can be obtained. 
Since the luminosity and color of the MSTO are sensitive to age as well as metallicity, we developed a new scheme, described in Paper~II, to estimate the distance modulus of the MSTO for a range of age and metallicity, based on the PARSEC theoretical isochrones \citep{Bressan+2012}. 
The resulting sample consists of MSTO stars, each with a typical uncertainty in the distance modulus of $\sigma_{\mu} \sim 0.43$, which corresponds to an approximate relative distance error of $\sigma_{D_{\odot}} / D_{\odot} \sim 0.2$, where $D_{\odot}$ is the distance of the individual MSTO star from the Sun.

This MSTO sample is used to derive the structural parameters of the stellar halo, fully taking into account the distance errors for the sample, the limited observational coverage of the sky, and the effect of the solar position relative to the Galactic Center.
Thus, we established a new scheme to obtain the structural parameters of the stellar halo from this limited input, as described in Paper~II. 
Application of this method to the full sample of MSTO stars in the range of $D_{\odot}=$ 13 to 50 kpc toward the North field, revealed that the number density profile of the stellar halo can be modeled by a spherical double power law, with an inner slope of $-3.28$, an outer slope of $-4.76$ and a break radius of 17.48~kpc. 
In the following analysis, we adopt this model for the background smooth stellar halo, and search for evidence of additional tidal debris at $D_{\odot} >$ 50 kpc, as predicted in the second-passage model for the orbit of the LMC.

In order to quantify the existence of substructure against the smooth halo components, we construct a mock color-magnitude diagram (CMD) in the plane of $(g-i)_{0}$ vs $i_{0}$.
We assume the MW's stellar halo is primarily composed of a stellar population with the age of 11.5~Gyr and ${\rm [M/H]} = -1.6$~dex, which approximate the mean values derived for our MSTO sample. 
Then, using the Kroupa IMF \citep{Kroupa+2001}, we can generate a mock Hertzsprung-Russell diagram. 
In order to obtain the mock observed CMD, we first generate a mock stellar halo following the spatial distribution based on the derived structure parameters for the background smooth halo model, and the stellar population is cross-matched with this spatial distribution.
Then, we divide the observed HSC magnitudes into 30 bins and compute the mean photometric uncertainty in each bin. 
The resulting relation between apparent magnitude ($m$) and uncertainty ($\sigma$) is interpolated using a nearest-neighbor scheme to construct a step-wise function $\sigma(m)$. 
We then generate mock observed magnitudes by adding Gaussian noise, with zero mean and a standard deviation given by $\sigma(m)$, to the modeled magnitudes, which are determined from the IMF, isochrones, and stellar positions.

%%%% Fig.3 %%%%%%%%%%%%%%%%%%%%%%%%%%%%
\begin{figure*}[ht!]
    \plotone{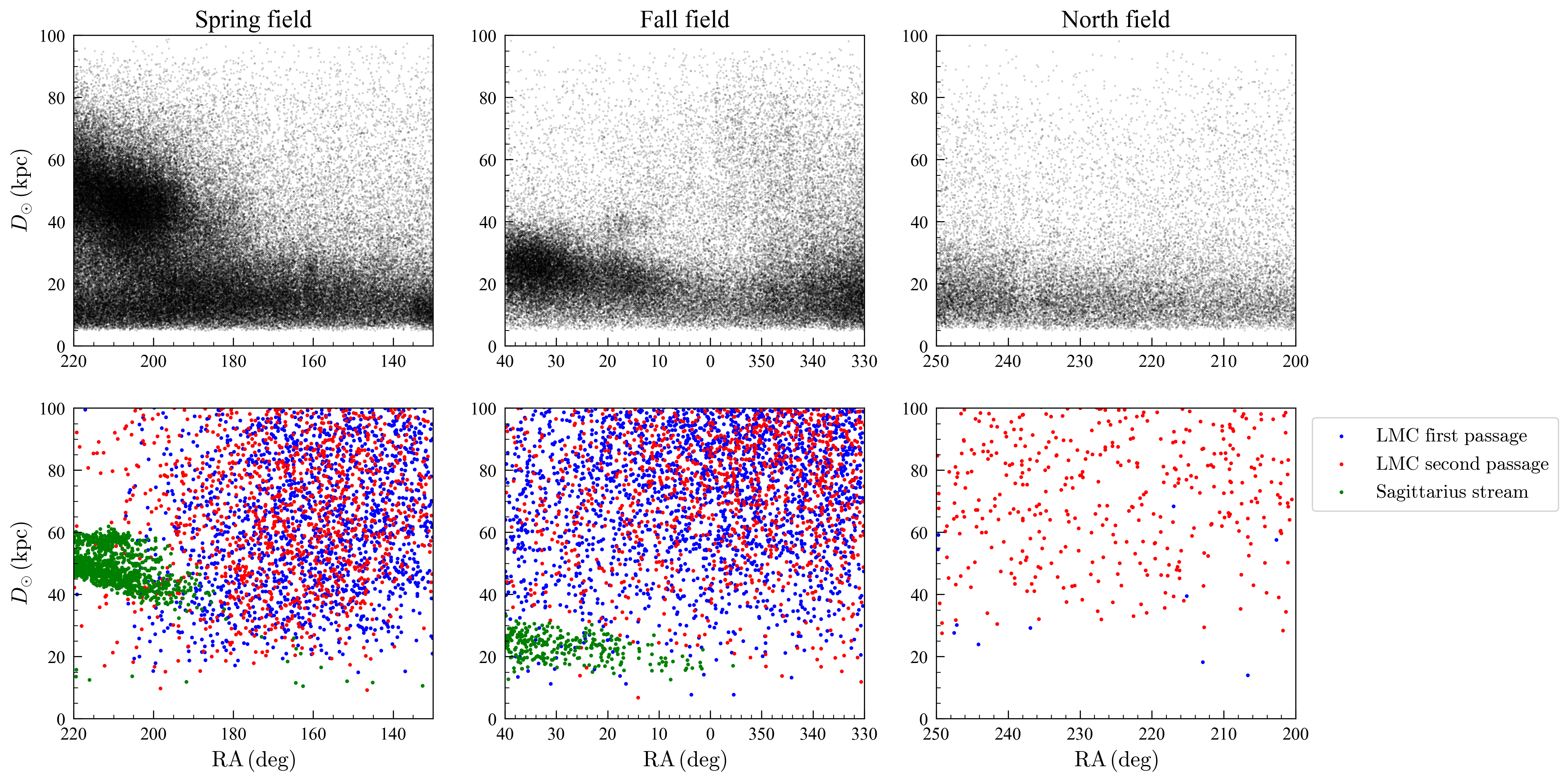}
    \caption{
    Comparison between the observed distribution of MSTO stars in the plane of right ascension vs heliocentric distance (upper panels) and the predicted tidal debris from simulations of the interactions between the MW and satellite galaxies (lower panels). 
    The green dots represent tidal debris from the Sagittarius dwarf, based on the simulation of \citet{Vasiliev+2021}. 
    The blue dots represent the predicted tidal debris from the LMC in the simulation of \citet{Vasiliev+2023}, under the assumption that the LMC is currently on its first passage (L2M10first model). 
    The red dots represent the tidal debris from the LMC in the simulation  of \citet{Vasiliev+2024}, under the assumption that the LMC is currently on its second passage (L2M10 model).
    It is apparent that the predictions for the  Sagittarius Stream are  consistent with the observational data in both the Spring and Fall Fields, and that both models of the orbit of the LMC predict debris in these fields. The rightmost lower panel shows very little predicted contribution from the recent passage of the LMC (first passage model; blue points) while there is tidal debris from the second passage model (red points), primarily beyond 50 kpc from the Sun, in a diffuse, featureless distribution. 
    }
    \label{fig3}
\end{figure*}
%%%%%%%%%%%%%%%%%%%%%%%%%%%%%%%%%%%%%%%

%%%%%%%%%%%%%%%%%%%%%%%%%%%%%%%%%%%%%%%%%%%%%%%%%%%%%%%%%%%%%%%%%%%%%%%%%%%%%%%%%%%%%%%%%%%%
% Section 3: Results
%%%%%%%%%%%%%%%%%%%%%%%%%%%%%%%%%%%%%%%%%%%%%%%%%%%%%%%%%%%%%%%%%%%%%%%%%%%%%%%%%%%%%%%%%%%%

\section{Results}

\subsection{Color Magnitude Diagrams}
Figure~\ref{fig2} shows the projected distribution of the halo stars in the North field (top panel) and their CMDs (bottom panels).
We also overplot the isochrone with age of 11.5~Gyr, ${\rm [M/H]} =-1.6$, and the distance from the Sun of $D_{\odot}=$~30~kpc, to guide the eye.  
As is seen, there exist clear signatures of substructure below the main sequence of this isochrone, at $i_{0}>$~22 mag, beyond $D_{\odot}=30$ kpc as indicated by the red arrow.
Interestingly, this structure can be detected across the extent of the North field as shown in the bottom panel of Figure~\ref{fig2}.
This implies that this substructure is widely dispersed, indeed spread over at least 100~$\mathrm{deg^2}$. 
As noted in Paper~I, there are only a few known stellar streams within a few kpc from the Sun in the area of sky covered by the North field \citep{Ibata+2019}, but the derived apocenter is within 30 kpc. 
Thus this diffuse substructure does not originate from nearby stellar streams.

\subsection{Comparison with the predicted tidal debris of the LMC}
We then compare the spatial distributions of the observed halo stars with those of the simulated tidal debris. 
In the top panels of Figure~\ref{fig3}, we show the distribution of the selected MSTO stars in the $(\mathrm{RA},D_\odot)$ plane, where the upper left, upper middle, and upper right panels correspond, respectively, to the Spring ($130^\circ<\mathrm{RA}<220^\circ$, $2^\circ<\mathrm{Dec}<4^\circ$), Fall ($330^\circ<\mathrm{RA}<400^\circ$, $-1^\circ<\mathrm{Dec}<1^\circ$), and North fields (cyan in Fig.~\ref{fig1}).
The lower panels of Figure~\ref{fig3} show the predicted particle positions of the tidal debris from the LMC based on the first-passage \citep{Vasiliev+2023} and second-passage models \citep{Vasiliev+2024} as the same models of Figure~\ref{fig1}, together with the distribution of particles from the simulated Sagittarius stream.
Note that since the distance uncertainty is $\sigma_{D_\odot}/D_\odot \sim 0.2$, the MSTO distribution is artificially broadened along the distance axis.  
Taking this broadening into account, the mean distance of the observed Sagittarius component in the upper panel is consistent with the models. 
It is important to note that only the second passage model for the LMC orbit deposits tidal debris in the North field, and much of this debris is at heliocentric distances beyond 50~kpc. 
Further, this debris is unclustered and spread across the area of the North Field. 
The predicted debris from the second passage model thus satisfies the distance and spatial uniformity inferred from Figure~\ref{fig2}.  

\subsection{Luminosity function}
\begin{figure}[ht!]
    \plotone{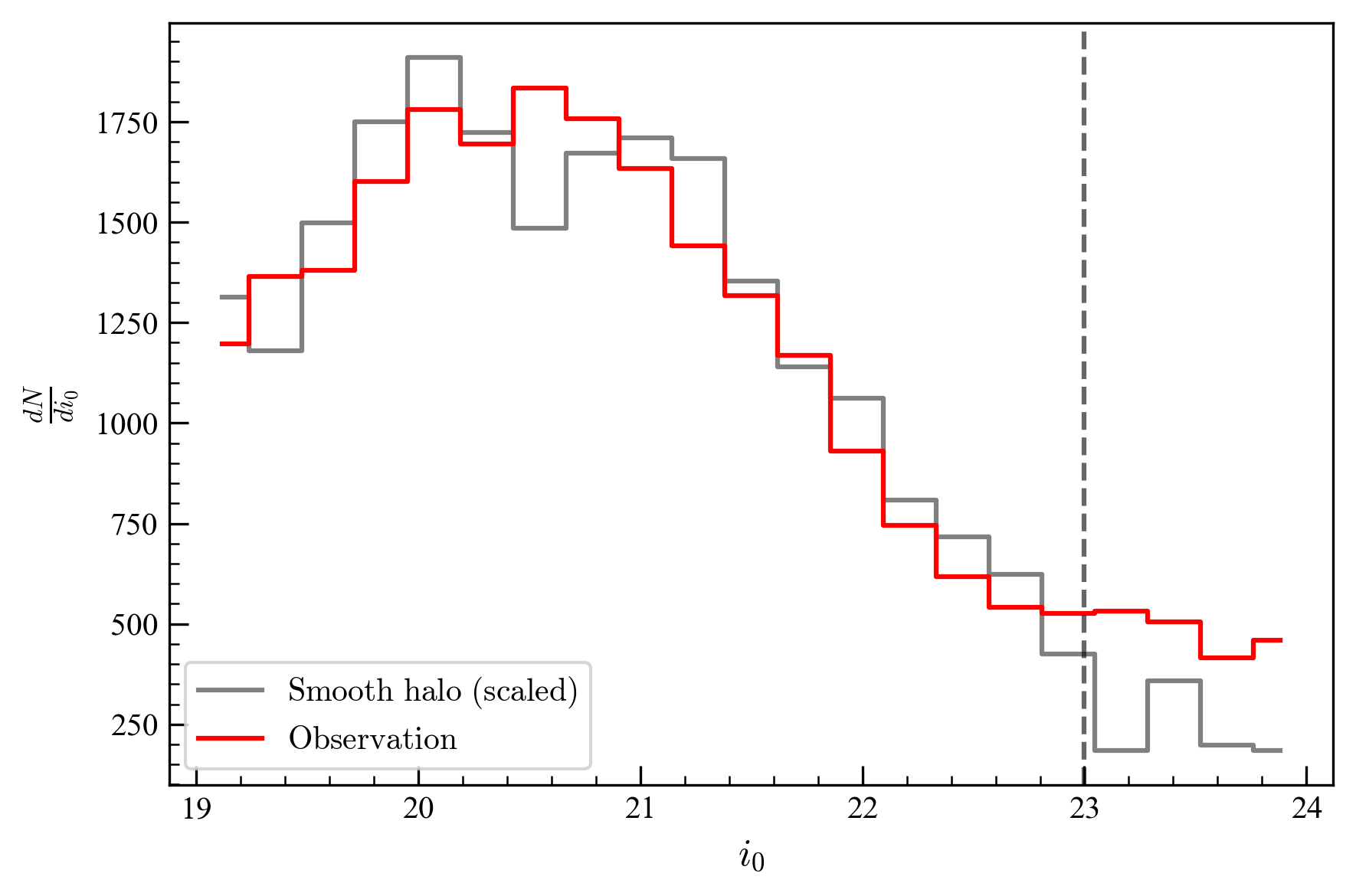}
    \caption{
    Luminosity functions for the smooth halo model and the observed halo. 
    The gray histogram shows the derived luminosity function for the smooth halo model toward the North field, adopting the best-fit parameters from Paper~II. 
    The red histogram shows the luminosity function for the observed halo stars. 
    The number counts are normalized in the $i$-band magnitude range of 19 to 21. 
    This figure shows an observed excess at faint magnitudes ($i_0 \gsim 23$) compared to the smooth halo component, indicating the presence of new substructure.
    }
    \label{fig4}
\end{figure}

Motivated by the possible association of the detected substructure with early debris from the LMC, we quantify to what extent the substructure is contributing to the North field. 
Figure~\ref{fig4} shows the comparison between the derived luminosity function of only the smooth model component of the stellar halo with that of the full observed distribution. 
We selected MSTO stars defined as $0.25 \le (g-i)_{0} \le 0.40$ and $19.0 \le i_{0} \le 24.0$ and normalized the counts using the interval of $19 \le i_{0} \le 21$~mag. 
We found that the excess fraction of stars over the smooth halo as defined by $f_{\rm excess} = \frac{N_{\rm obs} - N_{\rm halo}}{N_{\rm obs}}$ with $23.0 \le i_{0} \le 24.0$~mag is $0.48\pm0.06$, corresponding to the excess count of $907.9\pm123.2$, considering the 1-sigma uncertainty of the derived parameter for the smooth halo component.

%%%% Fig.4 %%%%%%%%%%%%%%%%%%%%%%%%%%%%
%%%%%%%%%%%%%%%%%%%%%%%%%%%%%%%%%%%%%%%

%%%%%%%%%%%%%%%%%%%%%%%%%%%%%%%%%%%%%%%%%%%%%%%%%%%%%%%%%%%%%%%%%%%%%%%%%%%%%%%%%%%%%%%%%%%%
% Section 4: Discussion and concluding remarks
%%%%%%%%%%%%%%%%%%%%%%%%%%%%%%%%%%%%%%%%%%%%%%%%%%%%%%%%%%%%%%%%%%%%%%%%%%%%%%%%%%%%%%%%%%%%

\section{Discussion and concluding remarks}
We identify a diffuse stellar excess in the outer halo of the MW, apparent in the CMD for stars beyond $\sim 30$~kpc from the Sun. 
This excess is traced by faint main-sequence stars and extends over at least $\sim100~\mathrm{deg}^{2}$ in the direction toward Bo\"otes (Figure~\ref{fig2}).
While the prominent outer halo substructures such as Sagittarius stream can be detected using MSTO stars, this substructure cannot be clealy seen (Figure~\ref{fig3}), indicating that this substructure is characterized by a very low surface brightness, making it most clearly detectable in the CMD and luminosity function rather than in direct spatial density maps.
When the smooth stellar halo component inferred within 50 kpc in Paper II, is extrapolated to larger Galactocentric distances, the observed luminosity functions in the color range $0.25 \leq (g-i)_0 \leq 0.40$ exhibit a excess fraction of stars of $\sim$50\%, corresponding to the excess count of $\sim$900 within $23.0 \leq i_0 \leq 24.0$ (Figure~\ref{fig4}).
To account for this excess, the presence of an additional diffuse outer halo component is required.

The fact that such a component has not been identified in previous studies is likely a direct consequence of low surface brightness. 
Most searches for substructure in the stellar halo employed intrinsically bright but rare tracer populations, such as blue horizontal branch (BHB) stars \citep{Amarante+2024}, RR Lyrae variables \citep{Sesar+2017}, or K-giants \citep{Conroy+2021}. 
While these tracers are well suited for identifying compact streams or high-contrast remnants of disrupted satellites, their low number densities severely limit the statistical sensitivity to diffuse, spatially extended structures. 
In contrast, the excess reported here becomes detectable only when large numbers of faint main-sequence stars are integrated over wide sky areas and examined in CMD and luminosity-function space, a regime that is accessible only with deep, wide-field imaging data such as those provided by Subaru/HSC.

Importantly, the same properties that made this structure difficult to detect also provide key constraints on its physical origin. 
Its broad spatial extent and low surface brightness argue against an origin as a dynamically cold stellar stream or the remnant of a recent, low-mass accretion event, which would be expected to produce sharper spatial features. 
Instead, these characteristics are more naturally explained by a component in dynamical relaxation, following substantial phase mixing.

Based on the results presented here, we propose three possible scenarios for the detected excess in the outer halo. 
First, in the LMC second-passage scenario, recent simulations \citep{Vasiliev+2024} suggest that predicted tidal debris from the LMC can be widely distributed at $D_{\odot}$ of $\sim100$ kpc.
The region probed by the North field overlaps closely with the predicted location of such debris, making this scenario geometrically plausible as shown in Figure~\ref{fig3}.
We also note that the higher total mass of the LMC ($2\times10^{11}~M_{\odot}$) assumed in the adopted simulations is consistent with recent estimates of a more massive central black hole inferred from high-velocity stars \citep{Han+2025}. 
However, without kinematic or chemical information, our imaging data cannot provide a definitive test of this hypothesis. 
Second, the excess could reflect a collective dynamical response of the MW dark matter halo to the LMC. 
While this feature have been reported using K-giant stars \citep{Conroy+2021}, subsequent analyses with larger BHB samples have failed to detect a statistically significant global response \citep{Amarante+2024}. 
If such a collective response were responsible, similar low-contrast features would also be expected in the other survey footprint of HSC-SSP, Spring and Fall fields, yet none are observed. 
Third, other scenarios cannot be excluded. 
The outer halo hosts various diffuse substructures with poorly understood origins such as the Outer Virgo Overdensity \citep{Sesar+2017} and the Pisces Overdensity \citep{Watkins+2009}, and the detected excess could alternatively represent debris from an unrelated accretion event rather than being associated with the LMC. 
Thus, while the LMC second-passage model is consistent with our data, it does not uniquely determine the origin of the excess.

With photometric information alone, the physical origin of the excess cannot be uniquely determined, and the present result should therefore be regarded as a candidate identification of an outer-halo substructure rather than a definitive attribution to a specific accretion event. 
Further progress will require combining deep, wide-area photometric surveys, such as UNIONS \citep{Gwyn+2025} and LSST \citep{Ivezic+2019}, with wide-field spectroscopic surveys such as PFS \citep{Takada+2014} and DESI \citep{Schlafly+2023}, which will provide line-of-sight velocities and chemical abundances for intrinsically brighter tracers at similar distances, even though existing facilities cannot reach the faint main-sequence stars analyzed here. 
These chemo-dynamical datasets will be essential for determining whether the newly identified component represents LMC tidal debris from a second passage or the remnant of another accretion event, contributing to a more complete understanding of the assembly history of the MW.

%%%%%%%%%%%%%%%%%%%%%%%%%%%%%%%%%%%%%%%%%%%%%%%%%%%%%%%%%%%%%%%%%%%%%%%%%%%%%%%%%%%%%%%%%%%%
% Acknowledgement
%%%%%%%%%%%%%%%%%%%%%%%%%%%%%%%%%%%%%%%%%%%%%%%%%%%%%%%%%%%%%%%%%%%%%%%%%%%%%%%%%%%%%%%%%%%%

\begin{acknowledgments}
This work is supported in part by JST SPRING (No. JPMJSP2114) and JST, the establishment of university fellowships towards the creation of science technology innovation (No. JPMJFS2102), and the Graduate Program on Physics for the Universe (GP-PU), Tohoku University for YS and JSPS Grant-in-Aid for Scientific Research and MEXT Grant-in-Aid for Scientific Research (No. 18H04334, 21H05448, 24K00669 and 25H00394) for MC. RFGW acknowledges support from Schmidt Sciences, through the generosity of Eric and Wendy Schmidt, by recommendation of the Schmidt Futures program. The HSC collaboration includes the astronomical communities of Japan and Taiwan and Princeton University.  The HSC instrumentation and software were developed by the National Astronomical Observatory of Japan (NAOJ), the Kavli Institute for the Physics and Mathematics of the Universe (Kavli IPMU), the University of Tokyo, the High Energy Accelerator Research Organization (KEK), the Academia Sinica Institute for Astronomy and Astrophysics in Taiwan (ASIAA), and Princeton University. Funding was contributed by the FIRST program from the Japanese Cabinet Office, the MEXT, JSPS, JST, the Toray Science  Foundation, NAOJ, Kavli IPMU, KEK, ASIAA, and Princeton University. This paper makes use of software developed for the Vera C. Rubin. We thank the LSST Project for making their code freely available. The Pan-STARRS1 (PS1) Surveys have been made possible through contributions of the Institute for Astronomy, the University of Hawaii, the Pan-STARRS Project Office, the Max-Planck Society and its participating institutes, the Max Planck Institute for Astronomy and the Max Planck Institute for Extraterrestrial Physics, The Johns Hopkins University, Durham University, the University of Edinburgh, Queen's University Belfast, the Harvard-Smithsonian Center for Astrophysics, the Las Cumbres Observatory Global Telescope Network Incorporated, the National Central University of Taiwan, the Space Telescope Science Institute, the National Aeronautics and Space Administration under Grant No. NNX08AR22G issued through the Planetary Science Division of the NASA Science Mission Directorate, the National Science Foundation under Grant No.AST-1238877, the University of Maryland, and Eotvos Lorand University (ELTE).
\end{acknowledgments}

%%%%%%%%%%%%%%%%%%%%%%%%%%%%%%%%%%%%%%%%%%%%%%%%%%%%%%%%%%%%%%%%%%%%%%%%%%%%%%%%%%%%%%%%%%%%
% Facilities and software
%%%%%%%%%%%%%%%%%%%%%%%%%%%%%%%%%%%%%%%%%%%%%%%%%%%%%%%%%%%%%%%%%%%%%%%%%%%%%%%%%%%%%%%%%%%%

\facilities{HSC~(Subaru)}

\software{
    Matplotlib \citep{Hunter+2007},
    Numpy \citep{Harris+2020},
    Scipy \citep{Virtanen+2020}
    Pandas \citep{Mckinney+2010}
    Astropy \citep{Astropy+2013, Astropy+2018, Astropy+2022}
}

%%%%%%%%%%%%%%%%%%%%%%%%%%%%%%%%%%%%%%%%%%%%%%%%%%%%%%%%%%%%%%%%%%%%%%%%%%%%%%%%%%%%%%%%%%%%
% Bibliography
%%%%%%%%%%%%%%%%%%%%%%%%%%%%%%%%%%%%%%%%%%%%%%%%%%%%%%%%%%%%%%%%%%%%%%%%%%%%%%%%%%%%%%%%%%%%

\bibliography{ref}{}
\bibliographystyle{aasjournalv7}

\end{document}